# Waterworlds Probably Do Not Experience Magmatic Outgassing

Joshua Krissansen-Totton[1,2,5], Max L. Galloway[1], Nicholas Wogan[2,3], Jasmeet K. Dhaliwal[4], and Jonathan J. Fortney[1]
[1] Department of Astronomy and Astrophysics, University of California, Santa Cruz, CA, USA; jkt@ucsc.edu
[2] NASA Nexus for Exoplanet System Science, Virtual Planetary Laboratory Team, University of Washington, Seattle, WA, USA
[3] Department of Earth and Space Sciences, University of Washington, Seattle, WA, USA
[4] Department of Earth and Planetary Sciences, University of California, Santa Cruz, CA, USA

## Abstract

Terrestrial planets with large water inventories are likely ubiquitous and will be among the first Earth-sized planets to be characterized with upcoming telescopes. It has previously been argued that waterworlds—particularly those possessing more than 1% $H_2O$—experience limited melt production and outgassing due to the immense pressure overburden of their overlying oceans, unless subject to high internal heating. But an additional, underappreciated obstacle to outgassing on waterworlds is the high solubility of volatiles in high-pressure melts. Here, we investigate this phenomenon and show that volatile solubilities in melts probably prevent almost all magmatic outgassing from waterworlds. Specifically, for Earth-like gravity and oceanic crust composition, oceans or water ice exceeding 10–100 km in depth (0.1–1 GPa) preclude the exsolution of volatiles from partial melt of silicates. This solubility limit compounds the pressure overburden effect as large surface oceans limit both melt production and degassing from any partial melt that is produced. We apply these calculations to Trappist-1 planets to show that, given current mass and radius constraints and implied surface water inventories, Trappist-1f and -1g are unlikely to experience volcanic degassing. While other mechanisms for interior-surface volatile exchange are not completely excluded, the suppression of magmatic outgassing simplifies the range of possible atmospheric evolution trajectories and has implications for interpretation of ostensible biosignature gases, which we illustrate with a coupled model of planetary interior–climate–atmosphere evolution.

*Unified Astronomy Thesaurus concepts:* Exoplanet atmospheres (487); Planetary atmospheres (1244); Planetary science (1255); Exoplanet evolution (491); Planetary interior (1248)

## 1. Introduction

Planet formation models (Raymond et al. 2004, 2013; Kite & Ford 2018) and exoplanet demographics suggest that water-rich terrestrial planets, or "waterworlds," are abundant (Zeng et al. 2019). Several candidate waterworlds will be amenable to atmospheric characterization with the James Webb Space Telescope (Krissansen-Totton et al. 2018; Fauchez et al. 2019; Lustig-Yaeger et al. 2019; Wunderlich et al. 2019). Indeed, preliminary mass constraints of the Trappist-1 planets suggested that they may all be volatile-rich (Grimm et al. 2018; Unterborn et al. 2018). Updated mass constraints show volatiles may not be required to explain their relatively low densities if iron fractions are also low, but the volatile enrichment remains a possibility for the outer planets: 1e, 1f, 1g, and 1h (Agol et al. 2021). Evaluating the capacity for volatile cycling between the atmosphere and interior will be crucial for interpreting future spectroscopic observations of waterworld atmospheres, and for interpreting any potential biosignature gases.

For the purposes of this paper, we define waterworlds as $<2\,M_{Earth}$ planets with water inventories that are easily sufficient to submerge all land (typically $>0.1$ wt% $H_2O$). While silicates and water become miscible under high pressures and temperatures such as those found within the interiors of sub-Neptunes (Nisr et al. 2020; Vazan et al. 2020), because we limit ourselves to Earth-sized waterworlds, we assume silicates and water separate into distinct layers during formation. Specifically, we consider planets where the majority of the planet's water inventory is degassed during magma ocean solidification and resides at the surface as steam, liquid water, or ice (e.g., Elkins-Tanton 2008; Hamano et al. 2013; Ikoma et al. 2018; Kite & Ford 2018).

The habitability and biosignature prospects of waterworlds are of particular interest because of the likely abundance of these planets and the potential for their near-term characterization (Cowan & Abbot 2014; Kitzmann et al. 2015; Noack et al. 2016; Kite & Ford 2018; Höning et al. 2019; Nakayama et al. 2019; Glaser et al. 2020; Hayworth & Foley 2020). Notable challenges to habitability include the formation of high-pressure ices that could impede the exchange of nutrients between the silicate interior and the surface (Noack et al. 2016; Journaux et al. 2020), the continuous drawdown of carbon dioxide via seafloor weathering decoupled from surface climate (Nakayama et al. 2019), and the low supply of bio-limiting phosphorus from seafloor weathering (Glaser et al. 2020). None of these potential challenges is insurmountable, however. Convection of high-pressure ices may ensure some mantle–surface exchange of nutrients (Choblet et al. 2017; Journaux et al. 2017; Kalousová et al. 2018), and habitable surface temperatures can potentially be maintained for several billion years for very large water inventories (Kite & Ford 2018). Additionally, the rates of phosphorus liberation on waterworlds may, in fact, be comparable to that of the Earth (Pasek et al. 2020; Syverson et al. 2020). Temperature dependent carbon cycle feedbacks on waterworlds may even be better suited for maintaining habitability than on planets with subaerial continents (Hayworth & Foley 2020).

It has been argued that terrestrial planets with very large surface water inventories ($>1$%) may not experience any volcanic activity or tectonics (Earth's oceans are 0.02% of planet mass). This is because the pressure overburden of surface water would act to increase the silicate solidus, which suppresses partial melting, thereby shutting down outgassing and tectonics for ocean depth pressures exceeding 1–3 GPa (Noack et al. 2016; Kite & Ford 2018). Kite & Ford (2018) explored the consequences of negligible volatile exchange between the interior and ocean in detail. They found that climate evolution is governed exclusively by the stellar evolution and initial carbon and dissolved cation inventories. Moreover, in many cases, habitable surface conditions can be maintained for several gigayears if initial endowments of volatiles and cations are serendipitous, i.e., if the atmosphere–ocean partitioning of carbon dioxide results in clement surface temperatures.

---
[5] NASA Sagan Fellow.

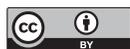

While zero volatile exchange with the interior is a conceptually useful endmember case, it has not been conclusively established that geologic activity and outgassing will shut down on all waterworlds rapidly after formation. For seafloor pressures <1–3 GPa, tectonics and geologic activity may of course continue (Höning et al. 2019; Nakayama et al. 2019). But even for large surface water inventories (>1–3 GPa seafloor pressures), interior heating from tidal interactions or large radiogenic inventories could conceivably yield super-adiabatic mantle plumes (e.g., Matyska & Yuen 2001; Barnes et al. 2013) that produce local or transient partial melting of the crust. Alternatively, if waterworld mantles become hydrated early in their evolution, perhaps due to efficient ocean–interior exchange when mantle potential temperatures are high, then this may suppress the silicate solidus, thereby decreasing the temperature required to melt, and hence allowing for increased partial melting (Katz & Cashman 2003). This scenario may offset the pressure overburden effect and permit geologic activity to persist.

There is another potential limit to post-magma ocean waterworld degassing that is not widely appreciated, however. The pressure overburden of a large surface ocean not only increases the mantle solidus, but also dramatically increases the solubility of volatiles in the melt phase. Consequently, even if melt production is large, volatiles are unlikely to exsolve and degas to the atmosphere–ocean reservoir; herein, this limitation is termed the "solubility limit to outgassing." For plausible melt compositions (i.e., maximum initial volatile content), this prevents the exsolution and degassing of volatiles prior to the suppression of partial melting that results from the pressure overburden. Although this effect has been described previously (Kite et al. 2009; Nakayama et al. 2019), it is typically omitted in studies of exoplanet outgassing. For example, Höning et al. (2019) assumed degassing of volatiles from partial melts according to mantle source composition, whereas Cowan & Abbot (2014) modeled the pressure overburden effect on outgassing and ingassing but not on melt solubility. In prior work that included a comprehensive set of exoplanet estimates for internal heatflow as an outgassing proxy, neither the pressure overburden effects on partial melting nor melt solubility were considered (Quick et al. 2020). Models of outgassing on non-waterworld stagnant lid planets (e.g., Noack et al. 2017; Dorn et al. 2018; Ortenzi et al. 2020) implicitly include solubility limits to outgassing by only allowing extrusive volcanism to contribute to atmospheric volatiles. For example, outgassed carbon fluxes are given by the product of melt production, carbon content in the melt, and the fixed proportion of all melt that erupts on the surface (typically around ∼10%). This approach cannot be easily applied to waterworlds because extrusive volcanism, including in submarine settings, may not contribute to volatile degassing under high pressures. Gaillard & Scaillet (2014) provided a theoretical framework for the pressure dependence of outgassing speciation, but only consider degassing pressures found in our solar system ($10^{-10}$ to 0.1 GPa) and not the pressures expected on ocean worlds (>0.1 GPa). Tosi et al. (2017) explicitly modeled the pressure overburden effect on outgassing using melt solubility expressions comparable to those adopted here, but they did not consider water-rich planets and so the effect is minor in that study.

A possible solubility limit to outgassing is discussed in Kite et al. (2009) and incorporated into a model of waterworld atmospheric evolution in Nakayama et al. (2019). Herein, we compare our calculations to these previous studies and generalize their conclusions to all waterworlds. This paper is structured as follows. First, we outline the theory behind solubility limits to outgassing, and demonstrate why waterworlds with >∼1 GPa seafloors are unlikely to degas volatiles regardless of mantle temperatures or volatile content. Note that these limits to outgassing apply only after accretionary magma oceans have solidified and catastrophic degassing is complete, and a liquid water ocean or ice layer has formed. Second, we apply these calculations to the Trappist-1 outer planets. Specifically, we use mass and radius constraints to infer their likely seafloor pressures, and we show that these planets are unlikely to degas volatiles if their low densities are attributable to a water-rich bulk composition. Next, the solubility limit to outgassing is put in the context of a systems evolution model for terrestrial planets to show the implications for geochemical cycling and climate evolution on Gyr timescales, again using Trappist-1 as an example. Finally, we examine our equilibrium assumptions and present arguments for why this is a reasonable representation of volatile loss over decompression paths of ascending magma.

## 2. Methods

Total outgassing fluxes from magmatic sources are determined by the integrated loss of volatiles over the decompression path from the mantle to the surface as magma ascends and cools (Edmonds & Woods 2018). When volatile-bearing magma moves toward the surface, the overburden pressure drops, and dissolved volatiles may reach saturation. Upon saturation, volatiles are exsolved from the magma into gas bubbles which may be released to the overlying atmosphere or ocean. Rather than explicitly model such decompression paths, we conservatively assume outgassing fluxes are determined by the melt-gas equilibrium upon magma emplacement (the validity of this assumption is explored in Section 4).

The solubilities of volatiles in magma can be represented by empirically derived relationships. For example, consider the Iacono-Marziano et al. (2012) solubility model for $CO_2$ and $H_2O$ in mafic melts:

$$\ln(x_{CO_2}) = x_{H_2O} d_{H_2O} + a_{CO_2} \ln(P_{CO_2}) + \frac{C_{CO_2} P}{T} + A_1 \quad (1)$$

$$\ln(x_{H_2O}) = a_{H_2O} \ln(P_{H_2O}) + \frac{C_{H_2O} P}{T} + A_2. \quad (2)$$

Here, $x_{CO_2}$ and $x_{H_2O}$ are mole fraction of $CO_2$ and $H_2O$ in the magma, respectively, where the carbon dioxide melt fraction includes both dissolved $CO_3^{2-}$ and $CO_2$. The variables $P_{CO_2}$ and $P_{H_2O}$ are the respective partial pressure of $CO_2$ and $H_2O$ in the magma in bars, $P$ is the total pressure in bars, and $T$ is the temperature in kelvin. The terms $d_{H_2O}$, $a_{CO_2}$, $C_{CO_2}$, $a_{H_2O}$, and $C_{H_2O}$ are empirical solubility constants from Iacono-Marziano et al. (2012), for an "anhydrous" case that is based on

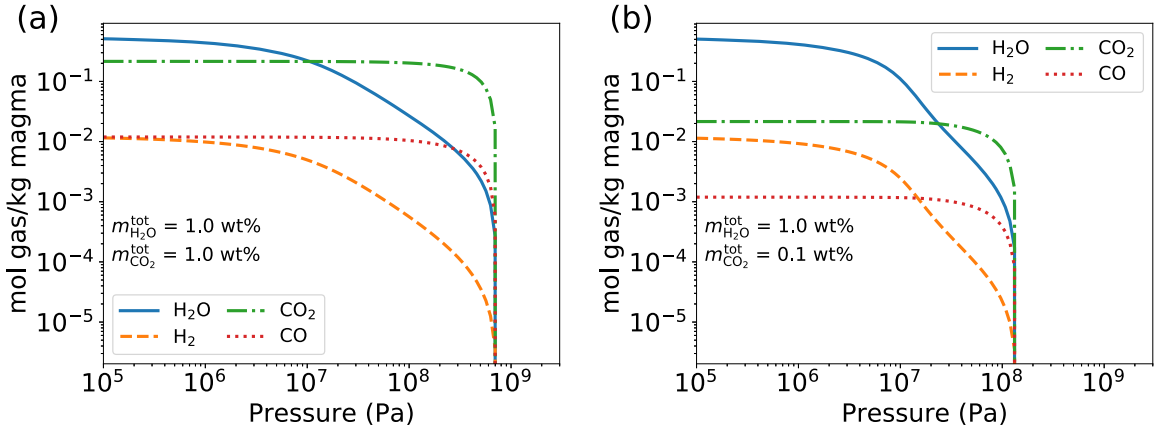

**Figure 1.** Outgassing solubility limit. Volatile abundances in mol kg$^{-1}$ of magmatic melt are plotted as a function of magma chamber pressure. (a) Melt is assumed to contain 1 wt% water and 1 wt% carbon dioxide by mass; (b) melt is assumed to contain 1 wt% water and 0.1 wt% carbon dioxide by mass. The melt is assumed to be buffered to fayalite–magnetite–quartz mineral redox buffer at 1473 K and equilibrium speciation of C–O–H-bearing species between the melt and gas phase are explicitly calculated, and we use the Iacono-Marziano et al. (2012) solubility model for H$_2$O and CO$_2$. In the case of the CO$_2$ rich melt (a), a pressure overburden more than 0.7 GPa will inhibit outgassing of all volatiles, whereas in the case of a lower CO$_2$ content melt (b), this cutoff is around 0.1 GPa. Full details of outgassing calculations are provided in Wogan et al. (2020).

**Table 1**
Constants and Coefficients for the Anhydrous Iacono-Marziano et al. (2012) Solubility Model

| Constant | Value |
|---|---|
| $d_{H_2O}$ | 2.3 |
| $a_{CO_2}$ | 1 |
| $C_{CO_2}$ | 0.14 |
| $A_1$ (Etna) | −0.42 |
| $A_1$ (MORB) | −0.87 |
| $a_{H_2O}$ | 0.54 |
| $C_{H_2O}$ | 0.02 |
| $A_2$ (Etna) | −2.60 |
| $A_2$ (MORB) | −2.55 |

Marrocchi & Toplis (2005). The values of these constants are given in Table 1. Finally, $A_1$ and $A_2$ are solubility parameters that depend on the major element composition of the magma, namely non-bridging oxygen content. Note that we have rearranged the solubility expressions in Iacono-Marziano et al. (2012) to produce Equations (1) and (2). Also, the solubility expression for CO$_2$ in Iacono-Marziano et al. (2012) states dissolved carbon as Ln[CO$_3^{2-}$]$^{ppm}$. However, the term should instead be Ln[CO$_2$]$^{ppm\,by\,weight}$ (K. Iacovino, S. Matthews, P. E. Wieser, G. M. Moore, & F. Begue 2020, in preparation).

We can use the solubility relationship for CO$_2$ (Equation (1)) to estimate the overburden pressure where degassing begins as melt rises to the surface. We consider CO$_2$ solubility because CO$_2$ is likely the first volatile (among C–H–O bearing species) to be exsolved from magma due to its low solubility (Holloway & Blank 1994). Making the substitution $P_{CO_2} = P_{degass} = P$ in Equation (1) gives

$$\ln(x_{CO_2}) = x_{H_2O} d_{H_2O} + a_{CO_2} \ln(P_{degas}) + \frac{C_{CO_2} P_{degas}}{T} + A_1. \quad (3)$$

Here, $P_{degas}$ is the overburden pressure where degassing begins or, equivalently, the maximum pressure at which magmatic degassing is possible. This equation can be solved for $P_{degas}$ as a function of the host rock composition.

It is important to note that a variety of empirically derived solubility relationships exist within the literature, and so the precise form of Equation (3) will depend on the assumed solubility model. In this paper, we use VESIcal (K. Iacovino et al. 2020, in preparation) to estimate degassing pressures with a variety of solubility models including MagmaSat (Ghiorso & Gualda 2015), VolatileCalc (Newman & Lowenstern 2002)—which is a simplification of Dixon (1997)—and the Iacono-Marziano et al. (2012) solubility model (see Figure A1(a) for a comparison of solubility relationships). For most calculations we use MagmaSat because it is valid for the largest range of pressures and temperatures (K. Iacovino et al. 2020, in preparation).

To estimate outgassed species and fluxes, we use the melt-gas equilibrium outgassing model described in Wogan et al. (2020). To briefly summarize, the model estimates the composition of gas bubbles suspended in magma just prior to their release into the overlying atmosphere or ocean. Gas bubble composition is computed by solving a system of equations including the Iacono–Marziano solubility relationships for H$_2$O and CO$_2$ (Equations (1) and (2)), gas-phase equilibrium relationships, and mass conservation of hydrogen and carbon. In this paper, we use the Wogan et al. (2020) model to illustrate the effect volatile solubility has on outgassing rate (Figure 1), and to calculate outgassing rates when simulating Trappist-1f's atmospheric evolution. This outgassing code uses values of $A_1$ and $A_2$ (Table 1) which are appropriate for the alkali–basaltic magma (based on data from Mt. Etna, Italy). Additionally, we consider values of $A_1$ and $A_2$ for average Mid-Ocean Ridge basalt (MORB) magma (Table 1). Only C–O–H-bearing volatiles are modeled, and so we ignore nitrogen and sulfur species (see below for further discussion of N and S outgassing). Each outgassing calculation requires the following inputs: the initial concentrations of H$_2$O and CO$_2$ in the melt before degassing occurs, temperature and pressure of degassing, and the redox state of the melt, which is a function of host rock oxygen fugacity.

To explore the implications of these outgassing calculations for the Trappist-1 system, we estimated the implied ocean depth and seafloor pressures for the Trappist-1 planets using mass and radius constraints (Agol et al. 2021) and a modified version of the publicly available interior model described in Zeng & Seager (2008) and Seager et al. (2007). A complete description of our approach is provided in Appendix A. To briefly summarize, mass and radius constraints are used to solve for iron core, silicate mantle, and water/ice mass fraction distributions, along with self-consistent pressure-depth profiles. A broad range of high molecular weight atmospheric heights and equations of state (EOSs) for iron and silicates are permitted.

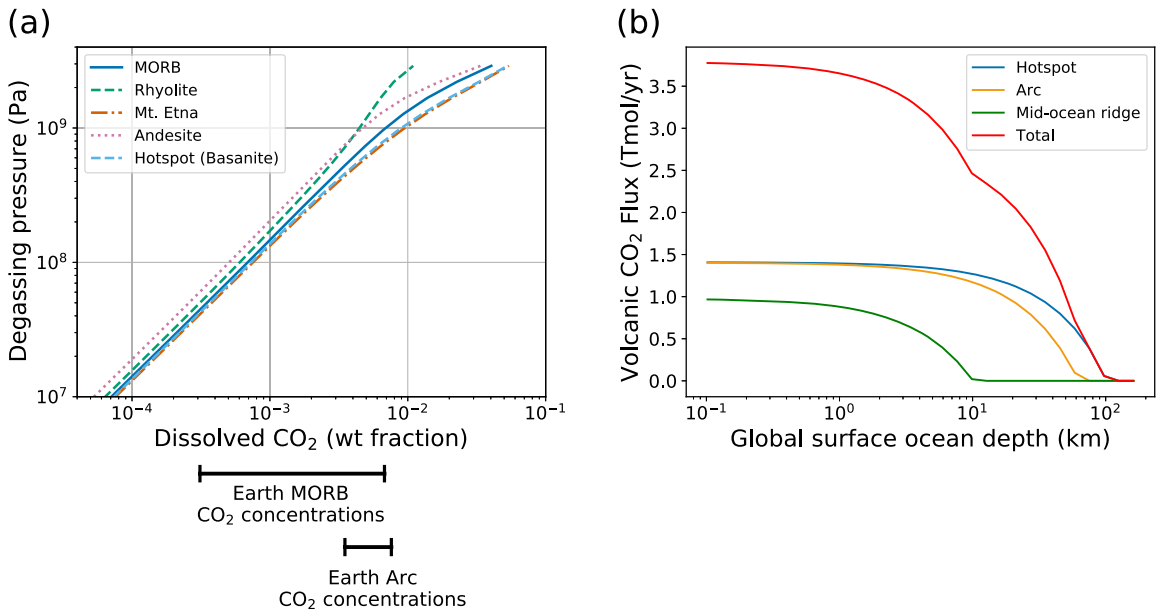

**Figure 2.** (a) Maximum degassing pressure as a function of wt fraction $CO_2$ in the melt for different bulk compositions. Compositions of Mid-Ocean Ridge basalt (MORB), rhyolite and andesite, Mt. Etna magma, and basanite are from Gale et al. (2013), Moore et al. (1998), Marrocchi & Toplis (2005), and Garcia et al. (1986), respectively. For comparison, we show the 95% confidence interval of Earth's observed MORB $CO_2$ concentrations (Le Voyer et al. 2019). For typical MORB $CO_2$ concentrations, no outgassing is expected for pressures exceeding ∼$10^3$ bar (0.1 GPa). For arc-like, $CO_2$ rich melts (0.35 to 0.76 wt%) (Fischer & Marty 2005), outgassing stops when ocean floor pressure exceeds $10^4$ bar (1 GPa). (b) Volcanic $CO_2$ fluxes as a function of globally averaged surface ocean depth, for different melt compositions assuming Earth-like gravity. We assume mid-ocean ridge magma has 0.1 wt% $CO_2$, arc magma has andesite composition with 0.6 wt% $CO_2$, and hotspot magma has basanite composition with 1 wt% $CO_2$ (Anderson & Poland 2017). Outgassing from $CO_2$-poor MORB ceases if pressure overburden exceeds ∼10 km of ocean, whereas arc-like and hotspot magmas cannot degas above ∼100 km of ocean. In both (a) and (b) we use the MagmaSat $CO_2$ solubility model with magma temperatures appropriate for each magma type, and assume the magma has no dissolved $H_2O$.

## 3. Results

Figure 1 illustrates the solubility limit to outgassing for two alkali basalts with different volatile compositions and melt redox state buffered to fayalite–magnetite–quartz. Outgassed volatile abundances are plotted as a function of the degassing pressure, assuming equilibrium partitioning between the melt and the gas phase. For melt compositions that are 1 wt% $H_2O$ and 1 wt% $CO_2$ (arc-like melts), no volatiles can be removed from the melt beyond ∼0.7 GPa (Figure 1(a)), whereas for melt compositions that are 1 wt% $H_2O$ and 0.1 wt% $CO_2$ (MORB-like melts), no volatiles can be removed from the melt beyond ∼0.1 GPa. These limits arise because gases are preferentially incorporated into the melt at high pressure, and beyond a certain threshold there are insufficient volatiles in the host rock to provide the equilibrium gas phase partial pressure, and the melt is undersaturated in both $CO_2$ and $H_2O$.

Figure 2(a) shows the $CO_2$ solubility relationships, or equivalently maximum degassing pressures as a function of dissolved melt $CO_2$. For typical MORB $CO_2$ melt fraction (∼0.1 wt%) (Le Voyer et al. 2019), degassing of $CO_2$ ceases for pressure overburdens exceeding ∼0.1 GPa. For dissolved $CO_2$ content more typical of arc volcanism (0.35–0.76 wt% $CO_2$)

(Fischer & Marty 2005), 1 GPa pressure overburden is required to prevent degassing. Note that the source rock bulk composition (e.g., felsic rhyolite, intermediate andesite, or more mafic MORB) has only a secondary effect on maximum degassing pressure compared to volatile content, i.e., for the same volatile content, the maximum degassing pressure varies by less than a factor of a few across all considered bulk rock compositions. Figure 2(b) shows $CO_2$ outgassing flux, assuming Earth-like crustal composition, as a function of global ocean depth (kilometers) added to an Earth-mass planet. The outgassing fluxes of $CO_2$ differ for hotspot, arc, and mid-ocean ridge volcanism due to differing host rock $CO_2$ content.

### 3.1. Comparison to Previous Results

Tajika & Matsui (1992) incorporated the pressure-dependence of $CO_2$ outgassing within an Earth system model, and this approach was adopted by Nakayama et al. (2019) to explore the atmospheric evolution of waterworlds. However, this outgassing formulation assumed that both melt and gas phase $CO_2$ degasses upon crystallization, and so even for 200 Earth ocean surface inventories (∼5 GPa seafloor), significant $CO_2$ outgassing was predicted to occur. This assumption may overestimate outgassing because melt phase volatiles will either be retained within the crystalline lattice or exsolved but trapped within melt inclusions upon solidification, as discussed below. Approximate estimates of the solubility limit to $CO_2$ degassing in Kite et al. (2009) are broadly consistent with what we illustrate in Figure 2. Specifically, Kite et al. (2009) argued that for 0.5 wt% $CO_2$ melt concentrations, 100 km of surface ocean would be sufficient to suppress $CO_2$ outgassing for an Earth-mass planet.

The solubility limit to outgassing implies that waterworlds ought to have extremely limited magmatic outgassing. However, other forms of volatile exchange may still occur. So long as there is new crustal production, serpentinization reactions may oxidize the crust and produce molecular hydrogen, although the efficiency of such reactions will depend on dissolved silica concentrations (Tutolo et al. 2020). While the possible tectonic regimes on waterworlds are uncertain, hydration and carbonatization of newly produced crust may, in principle, return volatiles to the mantle via subduction (Nakayama et al. 2019) or slab delamination (Foley & Smye 2018). Nakayama et al. (2019) developed this idea and hypothesized that if the water–rock interface is held at a constant temperature by the melting point of high-pressure ices, and if seafloor weathering reactions are purely temperature-dependent, then there will be no weathering thermostat to regulate waterworld climates. In this case, either outgassing fluxes exceed seafloor weathering rates and $CO_2$ accumulates in the atmosphere without bound, or—more likely if outgassing is suppressed by solubility limits—seafloor weathering sinks exceed $CO_2$ outgassing sources, and a runaway surface glaciation is inevitable.

The solubility limits described in this paper seem to make the runaway glaciations described in Nakayama et al. (2019) even more likely. It should be noted, however, that seafloor carbonatization rates will, to some degree, depend on the dissolved inorganic carbon content and pH of pore waters (Krissansen-Totton & Catling 2017 and references therein), as well as the carbon content of the undegassed oceanic crust (see Section 4). A broad range of dissolved inorganic carbon and ocean pH values are possible following accretion and magma ocean solidification (Kite & Ford 2018). Moreover, waterworlds with extremely carbon-rich mantles may degas if they yield $CO_2$ melt fractions more than a few wt% (see Section 4). Under stagnant lid regimes, the amount of fresh crust accessible to carbonatization reactions could be less. In general, such source and sink considerations highlight the need to place solubility limits to degassing in the broader context of atmospheric evolution models. Notably, solubility limits to outgassing dramatically simplify such models; this will potentially enable more straightforward, testable predictions for waterworld atmospheric compositions and climate. We explore the consequences of solubility-limited outgassing in Section 3.3.

### 3.2. Application to Trappist-1: Mass–Radius Constraints Are Suggestive of Waterworlds

Figure 3 shows the inferred pressure at the silicate–water interface based on mass and radius constraints in Agol et al. (2021). Separate probability distributions are shown assuming iron core fractions of 10%–20%, 20%–30%, 30%–40%, and 40%–50%. The top panel shows the results using the original EOSs for iron and silicates from Zeng & Seager (2008), while the bottom panel shows results using EOSs that are empirically adjusted to fit Earth's interior composition (see Appendix A). These two scenarios are considered endmembers for plausible interior structures. In either case, we find that the volatile-rich interiors with high-pressure silicate–water interfaces are probable. The outliers in the bottom panels (around 1 MPa) represent all model runs with negligible surface water.

For Trappist-1f, if iron core mass fractions from 10% to 50% are permitted and nominal EOSs are assumed, only 2.4% of solutions yield seafloor pressures less than 1 GPa (Figure 3(a)). Similarly, only 1.1% of solutions for Trappist-1g have seafloor pressures below 1 GPa. If EOSs are adjusted to better fit Earth's density structure (Figure A2), the probability of seafloor pressures below 1 GPa is still comparatively low: 19.6% for Trappist-1f and 11.8% for Trappist-1g (again for iron fractions from 10% to 50%). Results for Trappist-1e and Trappist-1h are similar and shown in Appendix B. These results are broadly consistent with those in Agol et al. (2021) and we provide an explicit comparison in Appendix B. It was necessary to repeat these internal structure calculations to self-consistently calculate seafloor pressure as a function of iron and water mass fractions.

Results from this study suggest seafloor pressures exceeding $\sim$1 GPa are sufficient to suppress magmatic outgassing in most cases. The expected seafloor pressures on the Trappist-1 outer planets therefore imply that there is a strong probability that none of these planets undergoes magmatic outgassing if their low densities are attributable to a high volatile content. A high volatile content may not be required if planetary iron fractions are low (<20%) compared to that of solar system objects (e.g., 33% for Earth, 26% for Mars, 29% for Venus, 68% for Mercury). Inner Trappist-1 planets are not considered because they are within the runaway greenhouse limit, and any surface water would become an extended steam atmosphere. However, thick steam atmospheres are unlikely for the Trappist-1 inner planets based on current mass–radius constraints (Turbet et al. 2019; Agol et al. 2021).

### 3.3. Application to Trappist-1: Implications for Atmospheric Evolution

To illustrate the implications of a solubility limit to outgassing for waterworld atmospheric evolution, we incorporated solubility-limited outgassing into a coupled model of planetary geochemical, thermal, and climate evolution. The model, which is described in full in Krissansen-Totton et al. (2021), considers terrestrial planet atmospheric evolution from a post-accretion magma ocean onward. C–O–H-bearing volatiles are partitioned between the magma ocean and the atmosphere assuming chemical equilibrium, and then following magma ocean solidification volatile exchange between the atmosphere and interior may occur via magmatic outgassing, serpentinization, carbonatization, and hydration reactions. A radiative convective climate model, parameterized mantle convection, and atmospheric escape are fully coupled to volatile evolution. Crucially, the outgassing of volatiles depends on melt production, mantle volatile content, and on the pressure–temperature-dependent equilibrium partitioning of volatiles between the melt and gas phase as described above and in Wogan et al. (2020). Moreover, the silicate solidus depends on both pressure overburden and the mantle water content (Katz & Cashman 2003). Modifications to the model described in Krissansen-Totton et al. (2021) are described in Appendix C.

According to mass–radius constraints (Figure 3), there is a strong possibility that the water mass fraction of Trappist-1f exceeds $\sim$1%, especially if the planetary iron fraction is greater than 20%. This implies the depth of surface water/ice on Trappist-1f likely exceeds $\sim$100 km. Figure 4 shows possible evolutionary trajectories of Trappist-1f from a magma ocean phase to present assuming the modern surface volatile inventory exceeds 1 GPa (ocean depth >$\sim$100 km).

According to this model, Trappist-1f ought to have experienced a $\sim$100 Myr magma ocean due to the extended pre-main sequence of Trappist-1 (Figure 4(b)); absorbed shortwave radiation exceeds the runaway greenhouse limit for the first ∼100 Myr of Trappist-1f's evolution (Figure 4(e)). During this long-duration magma ocean phase, loss of hydrogen to space oxidized the mantle (Figure 4(h)), consistent with previous models of magma ocean evolution on planets around M dwarfs (Schaefer et al. 2016; Wordsworth et al. 2018; Barth et al. 2020). Once absorbed shortwave radiation drops below the runaway greenhouse limit, the magma ocean freezes and a deep liquid water ocean condenses out of the steam atmosphere (Figure 4(d)). Despite the large pressure overburden from this surface ocean, melt production may persist to the present (Figure 4(g)). This is because crustal hydration reactions and subduction deliver water to the interior (Figure 4(i)), suppressing the silicate solidus and hence allowing for partial melting. However, despite high crustal production rates, there is no magmatic outgassing throughout Trappist-1f's evolution (Figures 4(f), (i)) due to the high solubility of volatiles at GPa pressures.

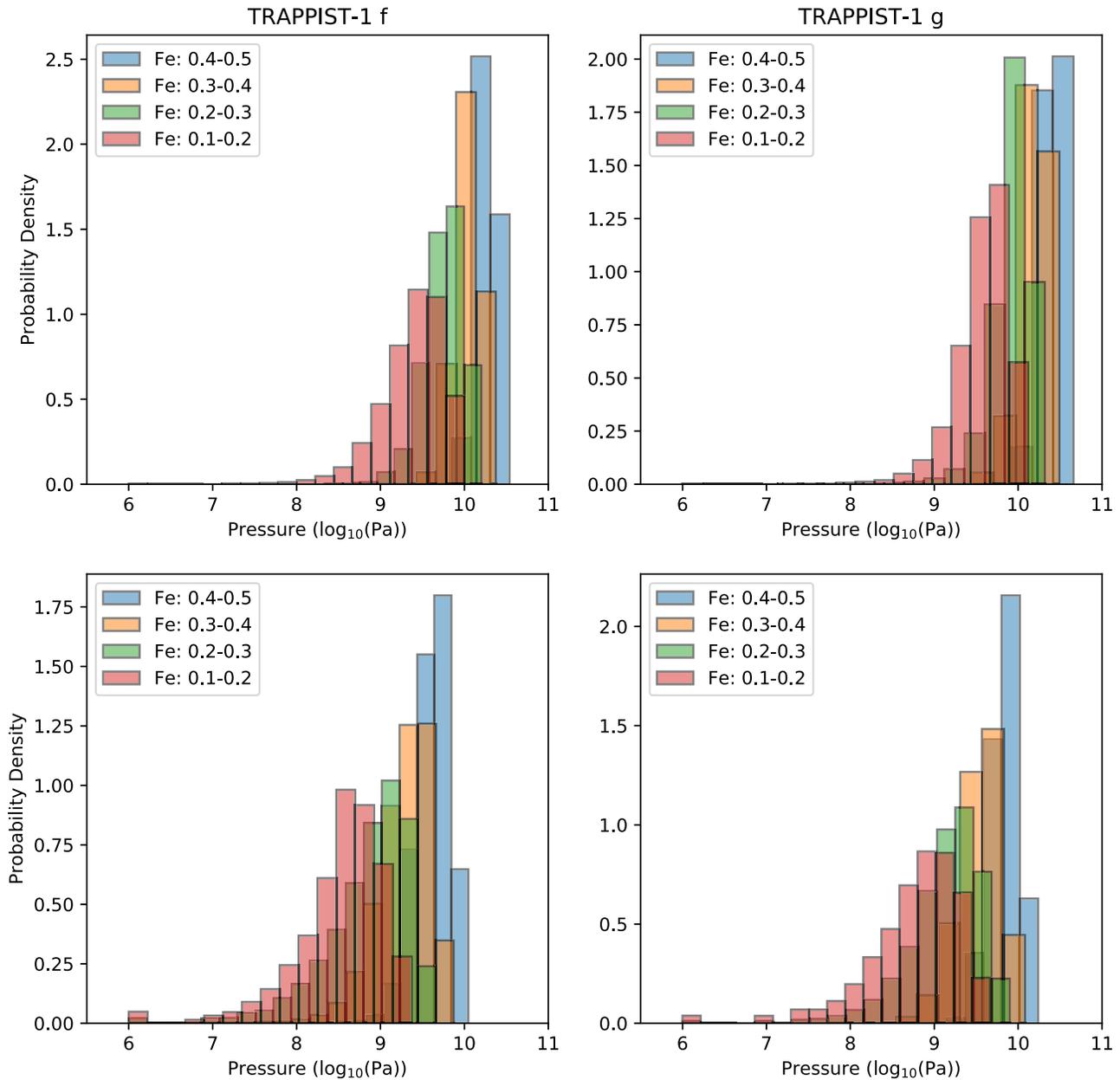

**Figure 3.** Pressure at the silicate–water interface for Trappist-1f and Trappist-1g. The top row shows results using the unmodified equations of state from Zeng & Seager (2008). The bottom row shows results using density profiles empirically modified to match Earth's pressure–density profile (see the Appendices). Probability distributions for seafloor pressure are obtained from solving for the full range of interior structures that fit observed mass and radius constraints (Agol et al. 2021). Red distributions represent probabilities assuming core iron fractions between 10%–20%, green distributions assume 20%–30% iron fractions, orange distributions assume 30%–40% iron fractions, and blue distributions assume 40%–50% iron by fraction mass. Water–silicate pressures below 1 MPa ($\log_{10}(Pa) < 6$) are plotted as 1 MPa; lower pressures indicate essentially zero surface water fraction. By comparison with Figure 2, we conclude that magmatic outgassing is unlikely on the outer Trappist planets due the suppression of exsolution from partial melts from to high-pressure overburden.

## 4. Discussion

To what extent is the simple melt-gas equilibrium partitioning approach described here a reasonable representation of waterworld degassing? Outgassing fluxes from magmatic sources are determined by the integrated loss of volatiles over the decompression path from the mantle to the surface as magma ascends and cools (Edmonds & Woods 2018). The decompression path reflects changes in volatile solubilities as magma ascends from its mantle source to the surface. Solubilities typically decrease with decreasing pressure, which leads to degassing of volatiles along the decompression path (so-called "first boiling"; Edmonds & Woods 2018). Additionally, as near-surface melt freezes and crystallizes, the melt phase may become enriched in volatiles that cannot be incorporated into a crystal lattice. In this crystallization scenario, if the melt is oversaturated in a volatile phase, it may lead to its exsolution into a fluid phase that coexists with the melt (so-called "second boiling"; Edmonds & Woods 2018); this process primarily depends on initial volatile inventories and their solubilities.

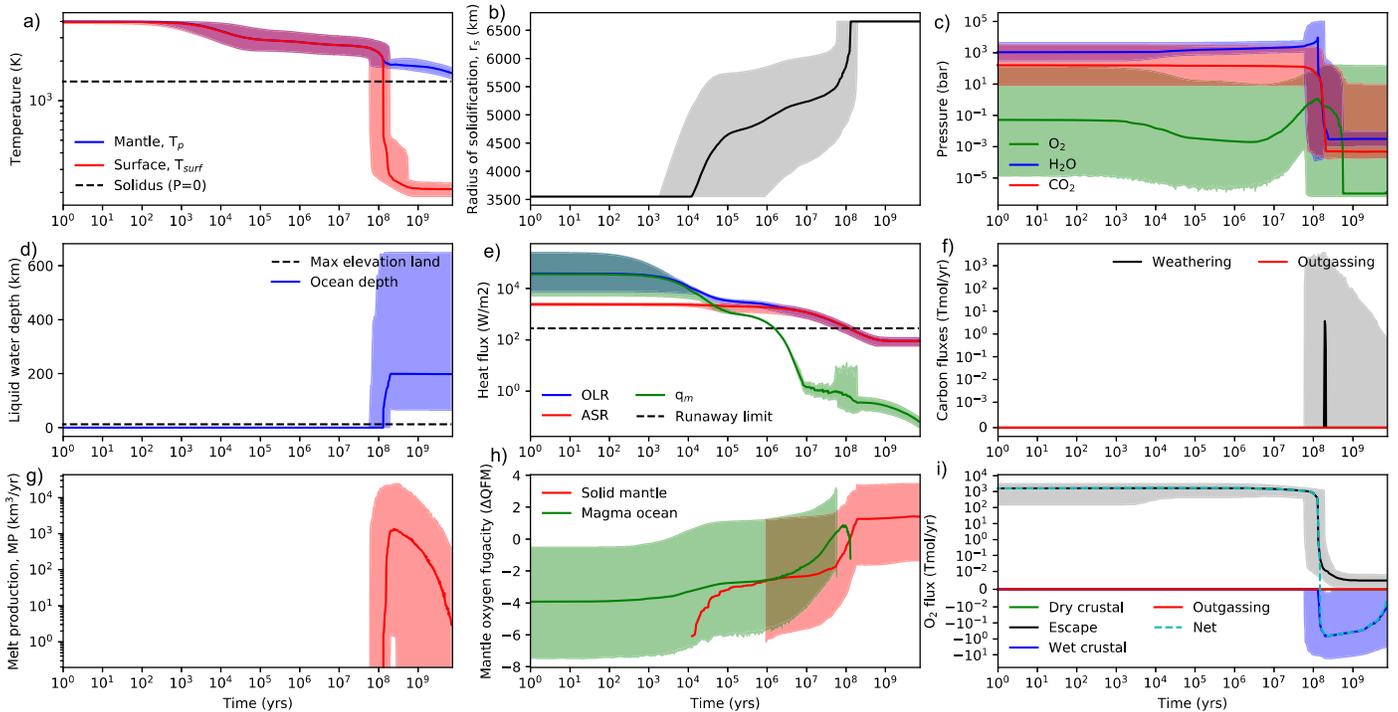

**Figure 4.** Trappist-1f's coupled redox–thermal–climate evolution. The model is applied from magma ocean to present with initial water inventories ranging from 30–300 Earth oceans (~1–10 GPa), and initial $CO_2$ inventories ranging from 20 to 6000 bar. The lines are median values and shaded regions denote 95% confidence intervals across all model runs. The coupled model includes (a) mantle and surface temperature evolution, (b) the solidification of the magma ocean from core to surface, (c) atmospheric evolution, (d) surface ocean condensation, (e) surface energy budget, (f) $CO_2$ outgassing and weathering fluxes, (g) crustal production, (h) mantle redox state, and (i) the surface redox budget. Despite the large pressure overburden from the surface ocean, crustal production may persist for billions of years due to mantle hydration suppressing the silicate solidus (g). Nonetheless, Trappist-1f does not experience magmatic outgassing due to the high solubility of volatiles in melt at GPa pressures (f), (i).

The solubility relationships adopted in this study approximate maximum degassing pressures by assuming gas-melt equilibrium. In real Earth magmas, volatile exsolution and degassing often initiate at pressures even lower than what would be predicted by these solubility relationships. This is because nucleating a gas bubble requires energy, and therefore bubbles only begin to form after the melt is over-saturated in a dissolved volatile species, meaning that melt volatile concentration is above its solubility at a given pressure (Burgisser & Degruyter 2015). In the case of homogeneous bubble nucleation, in which bubbles nucleate in a pure melt devoid of crystals, a melt-saturated rhyolite magma at 200 MPa will need to decompress down to 50–80 MPa before degassing begins (Burgisser & Degruyter 2015). Additionally, some MORBs are supersaturated in $CO_2$, unable to release all volatiles before the magma rapidly quenches (Bottinga & Javoy 1990).

Herein we argue that, regardless of the precise cooling trajectory, significant degassing of volatiles is unlikely at high seafloor pressures. Even on Earth, deep submarine extrusive eruptions occur at sufficiently high pressure, up to ~500–600 bar, to prevent exsolution of volatiles, particularly $H_2O$ (Wallace et al. 2015). Consequently, high volatile concentrations can be measured in volcanic glasses, which capture the melt composition, and form readily in submarine seetings because of rapid cooling at the lava–ocean interface. On waterworlds with >10 km oceans, both water and carbon-bearing species dissolved in extrusive melt would similarly not degas because of the large pressure overburden.

The same solubility limits to outgassing apply to magmas that cool more gradually. As melts crystallize, volatiles preferentially partition into the melt phase, and decreasing melt volumes lead to increasing dissolved volatile concentrations. These volatiles may eventually exsolve and become trapped in gas bubbles within melt inclusions (Moore et al. 2015). This volatile enrichment of the melt could allow degassing at pressures lower than what would be predicted by calculating solubility using source-rock volatile abundances. However, once trapped inside melt inclusions, subsequent bulk release of volatiles is unlikely. There is no evidence for post-entrapment diffusion of $CO_2$ into/out of melt inclusions (Esposito et al. 2014). Post-entrapment diffusion of H-bearing species may occur (Gaetani et al. 2012), but the comparatively high solubility of water makes this process less relevant at waterworld pressures.

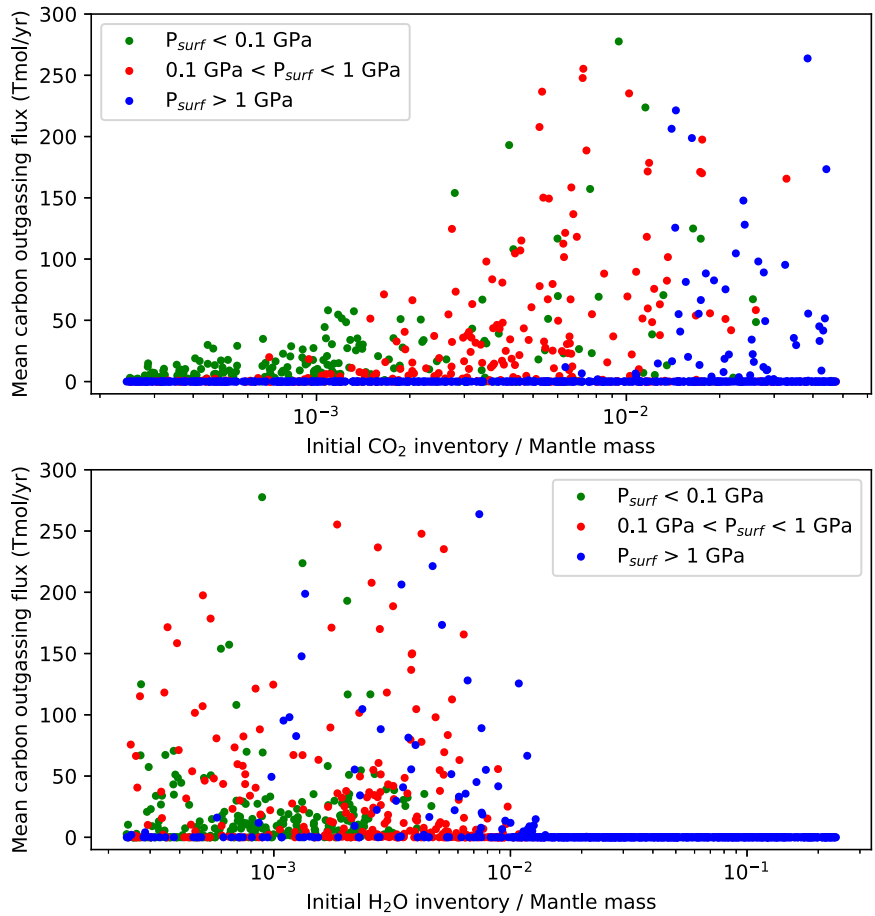

**Figure 5.** Effect of initial volatile inventories out the suppression of outgassing. Calculations from Figure 4 for Trappist-1f were repeated, but the initial carbon dioxide inventory range was extended from $10^{21}$ to $3 \times 10^{23}$ kg. Each dot represents an individual model run. Blue dots show model runs with >1 GPa final surface pressures, red dots show model runs with final surface pressures between 0.1 GPa and 1 GPa, and green dots show model runs with <0.1 GPa final surface pressure. (a) Average carbon dioxide outgassing fluxes plotted as a function of initial carbon dioxide inventory. We find that for carbon dioxide endowments that are a few wt% of mantle mass or greater, outgassing occurs in a small number of model runs. Outgassing only occurs for <1 wt% carbon content with surface pressures are lower than ∼1 GPa, whereas outgassing only occurs for <0.1 wt% carbon content when surface pressures are lower than ∼0.1 GPa. (b) Average carbon dioxide outgassing fluxes plotted as a function of initial water inventory. If the water inventory exceeds ∼1 wt% then outgassing does not occur, regardless of mantle carbon dioxide content.

In principle, changing pressure–temperature conditions due to metamorphism or later magmatic intrusions could result in re-equilibration and release of formerly entrapped volatiles, but again, the solubility limit to outgassing would apply. At most, these subsequent thermal metamorphism processes could ensure that the crystals and melts attain equilibrium with one another, if they had not before (i.e., all volatiles that can exsolve at a given pressure–temperature would exsolve). If thermal metamorphism were accompanied by uplift, then volatiles may be released because of changes in volatile solubilities at lower pressures. For waterworlds, however, this subsequent release will always be limited by the pressure overburden of the ocean above. Moreover, the temperature dependence of $CO_2$ and $H_2O$ solubility is negligible for our purposes (Figure A1(b)). Metamorphic volatile release via decarbonation of subducting slabs is similarly unlikely. Even in Earth's comparatively low-pressure subduction zones, direct decarbonation from carbon-rich metabasalts is small (Kerrick & Connolly 2001), and while arc magmas generated by the dehydration of subducting slabs could become enriched in carbon, they would be subject to pressure overburden at the ocean–silicate interface. Based on the discussion above, this study demonstrates that volatile retention equal to or greater than that predicted by melt solubility can be expected for magma on waterworlds.

Waterworlds with extremely $CO_2$-rich mantles could, in principle, generate sufficiently volatile-rich melts to exsolve carbon dioxide, even under high pressures (Figure 2). Figure 5 shows the results from a sensitivity test where Trappist-1f calculations from Figure 4 were repeated, but the initial carbon dioxide inventory range was extended from $10^{21}$ to $3 \times 10^{23}$ kg. Mean carbon dioxide outgassing fluxes are plotted as a function of initial carbon dioxide inventory for varying surface pressure ranges, and we find that for $CO_2$ endowments that are a few wt% of mantle mass or greater and surface pressure >1 GPa, outgassing occurs in a small number of model runs (Figure 5(a)). Note, however, that if the water inventory exceeds ∼1 wt% then outgassing does not occur, regardless of mantle carbon dioxide content (Figure 5(b)). Whether waterworlds with <1 wt% water are frequently endowed with a few wt% carbon dioxide depends on formation pathways (Raymond et al. 2004, 2013; Kite & Ford 2018). However, if mantle carbon content exceeds a few wt%, then geodynamics may differ dramatically to that of comparatively carbon-poor silicate mantles. Carbon-rich mantles may have higher viscosities and thermal conductivities, which would lead to a rapid shutdown of mantle convection and melt production (Unterborn et al. 2014). There may be a comparatively narrow range of carbon endowments that permit outgassing on waterworlds: too little carbon and carbon dioxide cannot exsolve from partial melts due to high-pressure overburdens, but too much carbon and the high mantle viscosity and conductivity preclude mantle convection and melt production.

The outgassing limits discussed in this paper apply to magmatic outgassing, namely the exsolution of volatiles from partial melts. Next, we consider the plausibility of other forms of interior-to-ocean volatile transfer. The solidification of magma on waterworlds will yield increasingly volatile-rich residual partial melts. An example of this melt enrichment is illustrated in Figure C1 using the thermodynamic melt model alphaMELTS (Ghiorso et al. 2002; Smith & Asimow 2005). Such volatiles may eventually exsolve and become trapped in inclusions or glasses upon solidification, with limited opportunities for subsequent release into the overlying ocean, as discussed above. However, for slow-cooling, intrusive melts with volatile-rich mantle source material, gravitational segregation may result in extremely volatile-rich, inviscid phases. The resulting mixture of C–O–H fluids, which would be miscible at waterworld pressures (Manning et al. 2013), could be transported to the ocean via hydrothermal circulation, crustal faulting, or metamorphism. The plausibility of generating large interior-to-ocean volatile fluxes via this mechanism is challenging to assess. In Earth's seafloor, even in old oceanic crust, the most extensive hydrothermal alteration and chemical exchange is limited to the top few hundred meters of pillow basalts; alteration declines with depth as pore spaces close and water/rock ratios decline (Staudigel 2003; Coogan & Gillis 2018). Under the high-pressure conditions of waterworlds, pore space closure would further limit diffusive and hydrothermal exchange. One opportunity for future research would be to simulate magma chamber cooling under waterworld conditions and subsequent reaction-transport of resultant volatile-rich phases.

While we focus on the C–O–H system, and particularly the outgassing of $CO_2$, the limits to outgassing described here may apply to other volatiles. In the case of sulfur, which experiences shallower degassing (Wallace et al. 2015 and references therein), the overburden pressure of waterworlds would be expected to limit outgassing. In contrast, nitrogen has comparatively low solubility in silicate melts, especially under oxidizing conditions (Libourel et al. 2003). However, Earth's (and Venus') nitrogen is depleted in combined mantle plus atmospheric reservoirs relative to other volatiles (Marty 2012), perhaps due to preferential partitioning of nitrogen into the metallic core during accretion (Roskosz et al. 2013). Subsequent mantle source rock nitrogen concentrations are thus considerably lower than that of $H_2O$ or $CO_2$, typically only a few ppm in MORB, at most (Busigny et al. 2005). Given such low host rock concentrations, nitrogen will be severely undersaturated in silicate melts at >0.1 GPa seafloor pressures (Libourel et al. 2003), and so nitrogen outgassing may also be suppressed on waterworlds, although further work quantifying nitrogen partitioning during planetary formation, as well as later geochemical cycling, is needed.

Outgassing pressures could be somewhat lower on waterworlds with significant seafloor topography, such as stagnant lid planets with large shield volcanoes. The maximum height mountains can achieve before flowing under their own weight is approximated by the following equation (Cowan & Abbot 2014):

$$h = 11.4 \left(\frac{g}{g_E}\right)^{-1} \quad (4)$$

where $h$ is maximum mountain elevation (in km), $g$ (m s$^{-2}$) is surface gravity, and $g_E = 9.8$ m s$^{-2}$ is Earth surface gravity. The maximum elevation variations on solar system planets are less than this upper limit. For Earth-sized planets, seafloor topography may therefore increase outgassing depths by a few kilometers, but this is a negligible effect for ~100 km deep oceans.

Our estimates of the outgassing pressure (silicate–water interface) for the Trappist-1 planets are limited by the accuracy and applicability of the assumed EOSs (see Appendix A for EOS details). The interior model does not consider the molten/solid state of the iron core or contributions from other siderophile elements such as carbon, both of which can impact density and therefore the inferred water mass fraction. Subtle changes in core density can result in large changes in the inferred pressure at the silicate–ocean interface. Similarly, by not taking into consideration a molten core, the model may be underestimating the minimum iron fraction and ocean depths for Trappist-1 planets. Our calculations only account for a radius decrease due to an ocean, not the additional pressure overburden from a dense atmosphere overlying surface water or ice. Therefore, the seafloor pressures may also be underestimated. In summary, the precise inferred silicate–water pressures are estimates, and could potentially be larger than what is suggested by the probability distributions shown in Figure 3. Such a scenario would imply the outer Trappist-1 planets are even less likely to undergo magmatic outgassing.

*4.1. Implications for Waterworld Climate Evolution and Redox Evolution*

The suppression of outgassing shown in Figure 4 often results in the secular drawdown of carbon dioxide and surface glaciation, similar to that described in Nakayama et al. (2019). Moreover, unlike Nakayama et al. (2019), our model of seafloor weathering accounts for the carbon dioxide dependence of seafloor weathering as surface reservoirs are depleted (Appendix C). Even with this supply limit to seafloor weathering, glaciation occurs eventually in most model runs because carbon dioxide replenishment via outgassing is completely suppressed by the pressure overburden. However, this treatment of carbon sinks does not account for possible limits to weathering due to the high carbon dioxide content of undegassed basalt. On Earth, unaltered MORB is comparatively carbon-poor (0.12 wt% Staudigel 2003), and so the release of cations from weathering of basaltic crust and subsequent precipitation of carbonates results in the net transfer of carbon from seawater to the crust. On waterworlds, however, the unaltered oceanic crust could be extremely carbon dioxide rich due to pressure overburden suppression of outgassing. Dissolution of oceanic crust via reactions with hydrothermal fluids would release both carbon and alkalinity into the pore space, potentially inhibiting the net uptake of oceanic carbon. Assessing the likelihood of waterworld glaciation would thus require further modeling of carbon-rich crustal dissolution and carbonate precipitation under high pressure conditions. In any case, temperate ocean surface conditions are still permitted in our Figure 4, specifically for model runs where melt production is negligible. This scenario can occur for very large water inventories and/or low radiogenic inventories where the shutdown of melt production halts the drawdown of carbon dioxide via carbonatization reactions. In such cases, temperate surface conditions may persist for several gigayears, as described in Kite & Ford (2018).

Solubility limits to outgassing also have potential implications for atmospheric redox evolution and biosignature false positives. For waterworlds that accumulate oxygen-rich atmospheres due to efficient hydrogen escape during their early evolution (Luger & Barnes 2015; Wordsworth et al. 2018; Barth et al. 2020), the absence of magmatic outgassing may hamper the post-magma–ocean drawdown of abiotic oxygen via reactions with degassed volatiles (e.g., $H_2$, CO, $CH_4$). This occurs in some model outputs in Figure 4(c). However, the absence of magmatic outgassing does not preclude oxygen drawdown via $H_2$ produced from serpentinization reactions, and so oxygen accumulation is not inevitable and depends on the balance between oxygen production via diffusion-limited hydrogen escape and oxygen consumption via hydration reactions (Figure 4(h)). The outputs shown in Figure 4 are purely illustrative, and a more comprehensive study of oxygen false positives on waterworlds will be a topic of future coupled-model analyses. The suppression of magmatic outgassing also strengthens the case for methane as a biosignature on terrestrial planets. Although high pressures may favor methane production, the high solubility of volatiles in partial melts makes large magmatic fluxes of methane on waterworlds unlikely (Wogan et al. 2020).

## 5. Conclusions

1. Terrestrial planets with large water inventories (>1% $H_2O$ bulk composition) are unlikely to experience magmatic outgassing because of the large pressure due to overlying water/ice. The threshold overburden pressure is around 0.1–1 GPa, assuming magmas on other worlds have comparable $CO_2$ concentrations to magma on Earth. This regime increases the solubility of C–O–H volatiles in mantle partial melts, severely limiting degassing from magmas and lavas.
2. If the apparent low densities of the Trappist-1 outer planets are attributable to a large water mass fraction, as opposed to a low iron content, then these planets have high overburden pressures that preclude magmatic outgassing.
3. Other forms of volatile exchange between the surface and mantle, such as carbonatization reactions and hydrothermal alteration, may still occur as long as new crust is being produced. However, the lack of volcanic degassing simplifies the atmospheric evolution for waterworlds and has implications for atmospheric redox and climate evolution. In particular, abiotic oxygen accumulation and glaciated climates are potentially more likely on waterworlds.

We thank Dennis Höning, Kayla Iacovino, Penny Wieser, and Quentin Williams for their insightful comments and helpful suggestions. J.K.T. was supported by NASA through the NASA Hubble Fellowship grant HF2-51437 awarded by the Space Telescope Science Institute, which is operated by the Association of Universities for Research in Astronomy, Inc., for NASA, under contract NAS5-26555. We also thank NASA's Virtual Planetary Laboratory (grant 80NSSC18K0829) for support.

## Appendix A
## Solubility Relationships and Method for Calculating Ocean Depth and Pressure Overburden

Figure A1 compares empirical solubility relationships and illustrates the temperature dependence of carbon dioxide solubility.

We adapted the publicly available code from Zeng & Seager (2008) and Seager et al. (2007) to calculate planetary interior composition for the Trappist-1 planets. The MATLAB code, ExoterDE, assumes a three-component interior composition including an iron core, a silicate mantle, and a water-ice outer layer. Planetary mass and radius are specified as inputs, and the optimization function ode45 is used to calculate density and pressure as a function of radius as well as iron, silicate, and water mass fractions. Solutions are degenerate, and so for any given mass and radius, there is a range of permitted iron–silicate–water fractions. EOSs for iron, silicate, and water/ice are described in Zeng & Seager (2008) (although see modifications below).

The code was adapted to perform Monte Carlo calculations. Observed mass–radius distributions were resampled 5000 times and interior composition was repeatedly calculated to obtain probability distributions for Trappist-1 interior structures. We also modified the code to allow for an atmosphere (with negligible mass) anywhere between 0 and 100 km in height, sampled uniformly. More extended $H_2$-rich atmospheres have been largely ruled out by transit observations (De Wit et al. 2018). More extended high molecular mass atmospheres are unlikely given that Trappist-1e, -1f, -1g, and -1h lie beyond the runaway greenhouse limit. Our modified version of the code is available on the Github of the lead author.

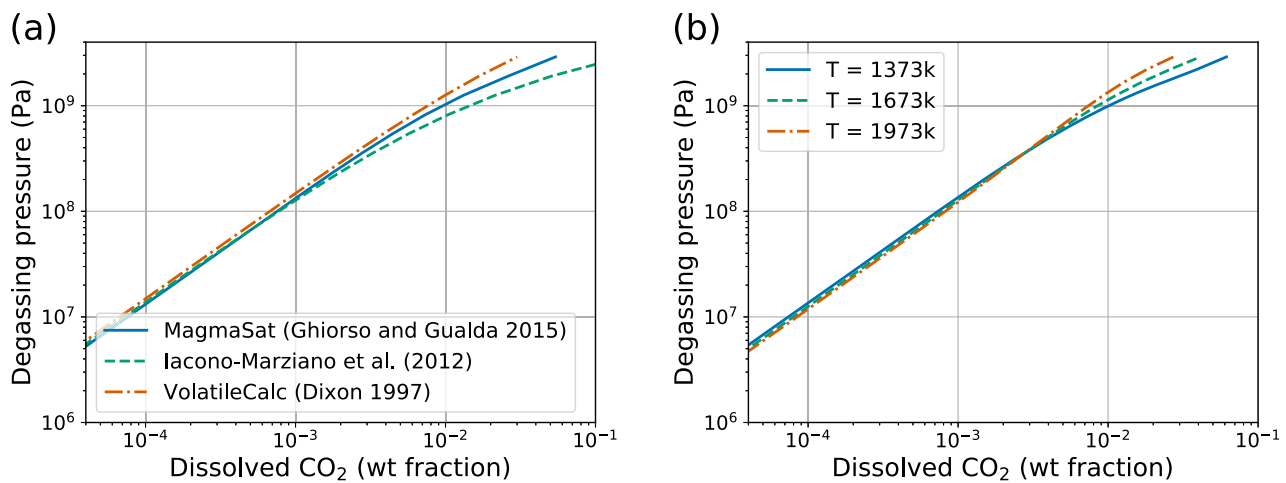

**Figure A1.** (a) Comparison of different empirical $CO_2$ solubility relationships and (b) temperature dependence of $CO_2$ solubility; (a) uses a magma temperature of 1473 K, (b) uses the MagmaSat $CO_2$ solubility model, and both (a) and (b) use Mt. Etna's magma composition (Iacono-Marziano et al. 2012). The differences between solubility models and the temperature dependence of solubility are minor for our purposes.

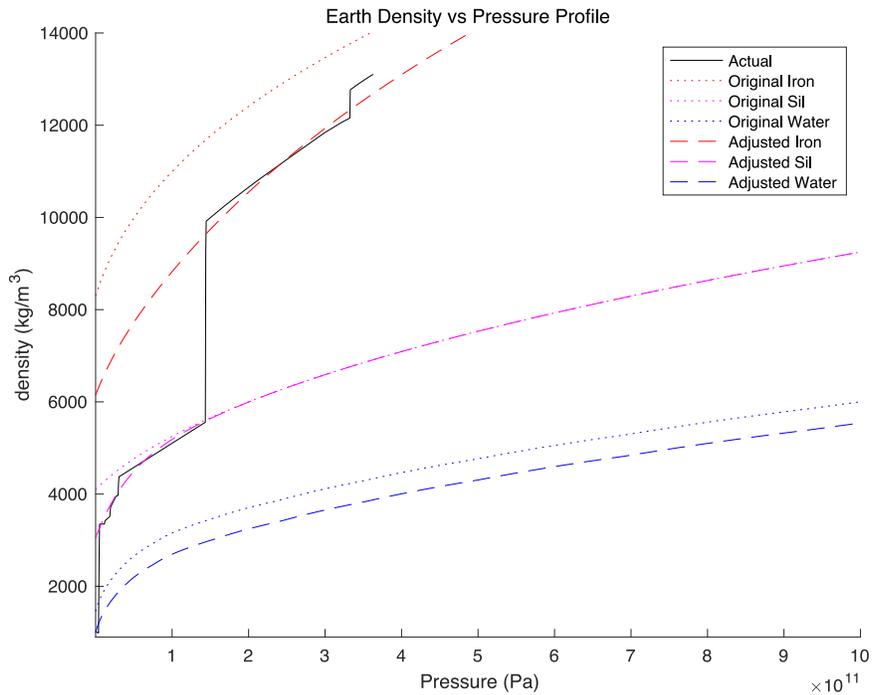

**Figure A2.** Comparison between Preliminary Reference Earth Model (solid black line), nominal density functions (dotted lines), and density functions empirically adjusted to fit Earth's density profile (dashed lines).

Two different density functions for the iron core, silicate mantle, and water/ice were used for mass–radius calculations. The first are the original density functions used in Zeng & Seager (2008): at low pressures most relevant to this study ($P < \sim 200$ GPa), experimental fits to the Vinet EOS or the Birch–Murnagham EOS are adopted. Higher pressures ($P > \sim 10^4$ GPa) assume the Thomas–Fermi–Dirac (TFD) EOS. Intermediate pressures for iron and silicate use the Vinet EOS until it intersects the TFD EOS. For water, the Birch–Murnagham EOS is used until 44.3 GPa, then theoretical data from density functional theory until it agrees with the TFD EOS at 7686 GPa. The assumed EOS and resulting density functions do not exactly reproduce Earth's internal structure, as shown in Figure A2, which compares density as a function of pressure for iron, silicate, and water model components to that of the Preliminary Reference Earth Model (Dziewonski & Anderson 1981). To fit the Earth's mass and radius with nominal density functions, a 2% surface water mass fraction is required, which is much larger than the true value of around 0.02%. Because of this, the original density profiles significantly overestimate the water composition and, therefore, overestimate the silicate–ocean interface pressure. Therefore, we made an empirical adjustment to better fit Earth's silicate and iron densities, which is also shown in Figure A2.

The Trappist-1 results using both the nominal density profiles and the empirically adjusted density profiles are both broadly consistent with internal structure estimates in Agol et al. (2021), as shown in Figure A3. The adjusted profiles tend to underestimate the water fractions reported by Agol et al. while the original density profiles slightly overestimate the water fractions. Neither set of results significantly deviates from those of Agol et al. The differences that are present are more significant for the higher iron fractions, indicating that the differences are likely due to different assumptions about core densities.

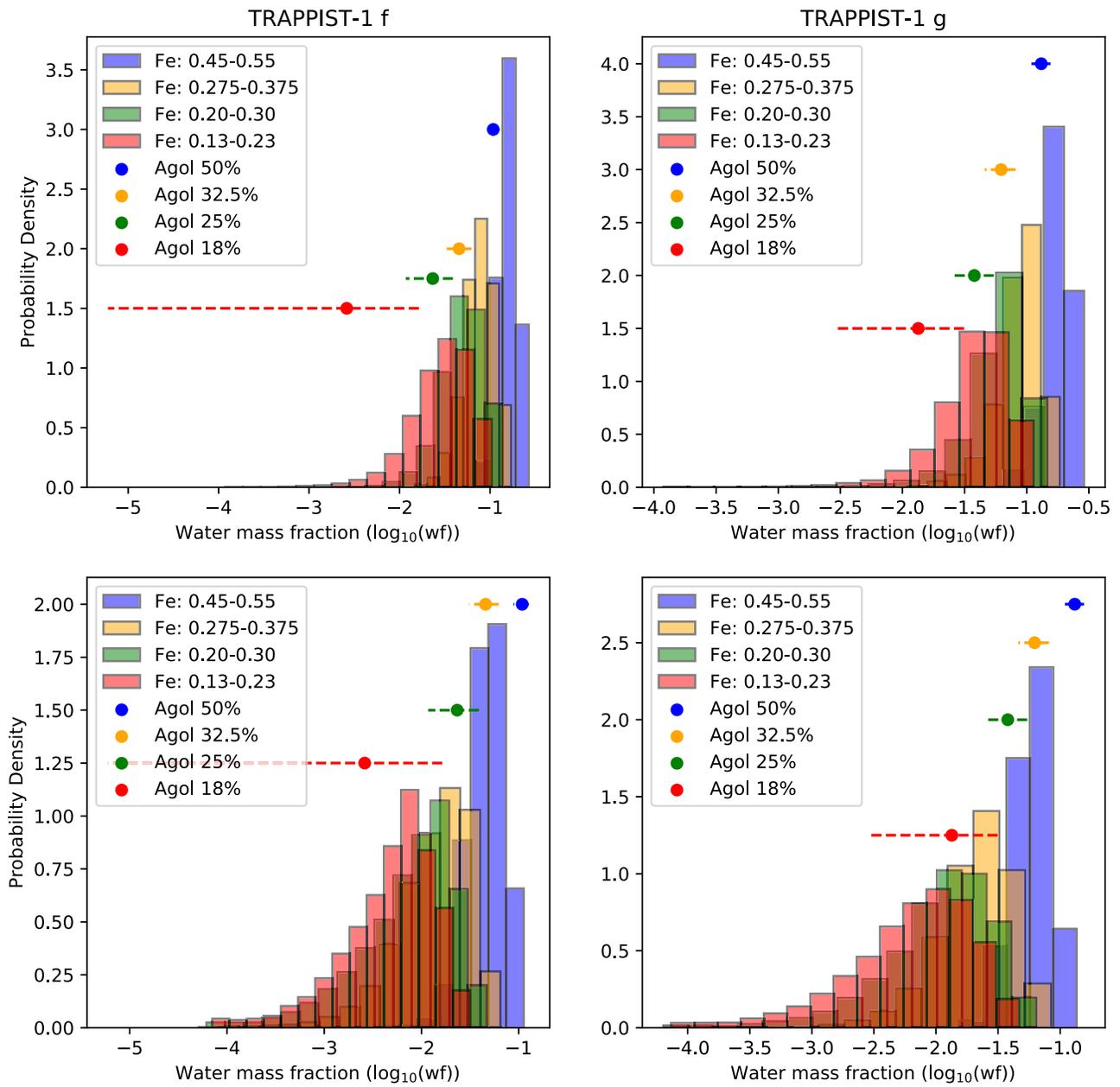

**Figure A3.** Water mass fractions for Trappist-1f and Trappist-1g compared to Agol et al. (2021). The top row plots the results using nominal density profiles and the bottom row plots the adjusted density profile results. The dots and dashed lines represent the $1\sigma$ uncertainty from Agol et al. while our results are shown using histograms. Results are in overall good agreement, although the nominal density profiles slightly overestimate water fraction whereas the adjusted density profiles underestimate the water fractions compared to Agol et al. Possible explanations for the slight discrepancies are discussed in the text.

## Appendix B
## Results for Trappist-1e and Trappist-1h

Equivalent calculations for Trappist-1e and 1h are shown in Figures B1 and B2.

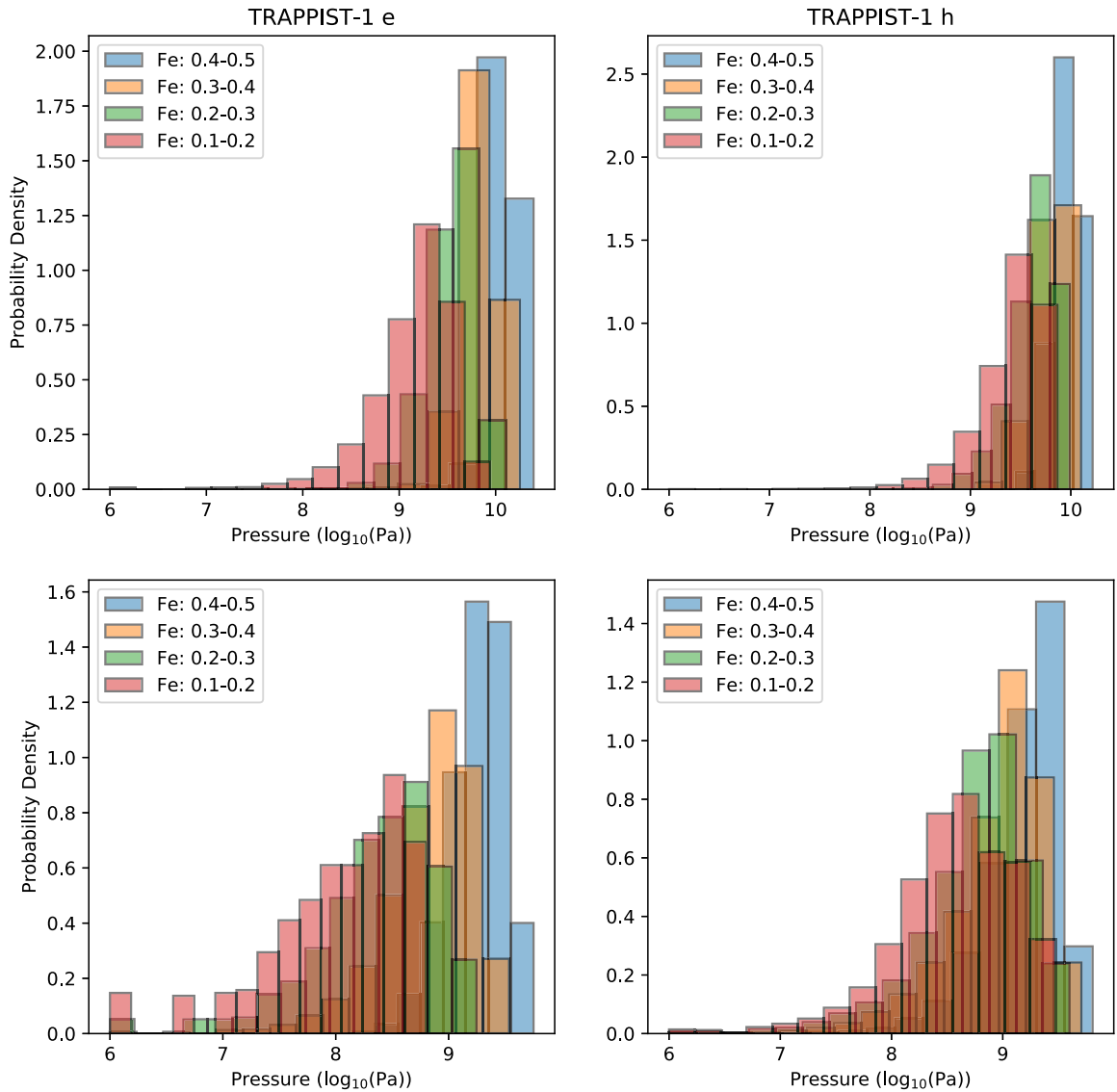

**Figure B1.** Identical to Figure 3 in the main text except analyses were repeated for Trappist-1e (left) and Trappist-1h (right). Volatile-rich compositions with high pressures at the water–silicate interface are probable for both planets, although 1e is less likely to require a large volatile fraction.

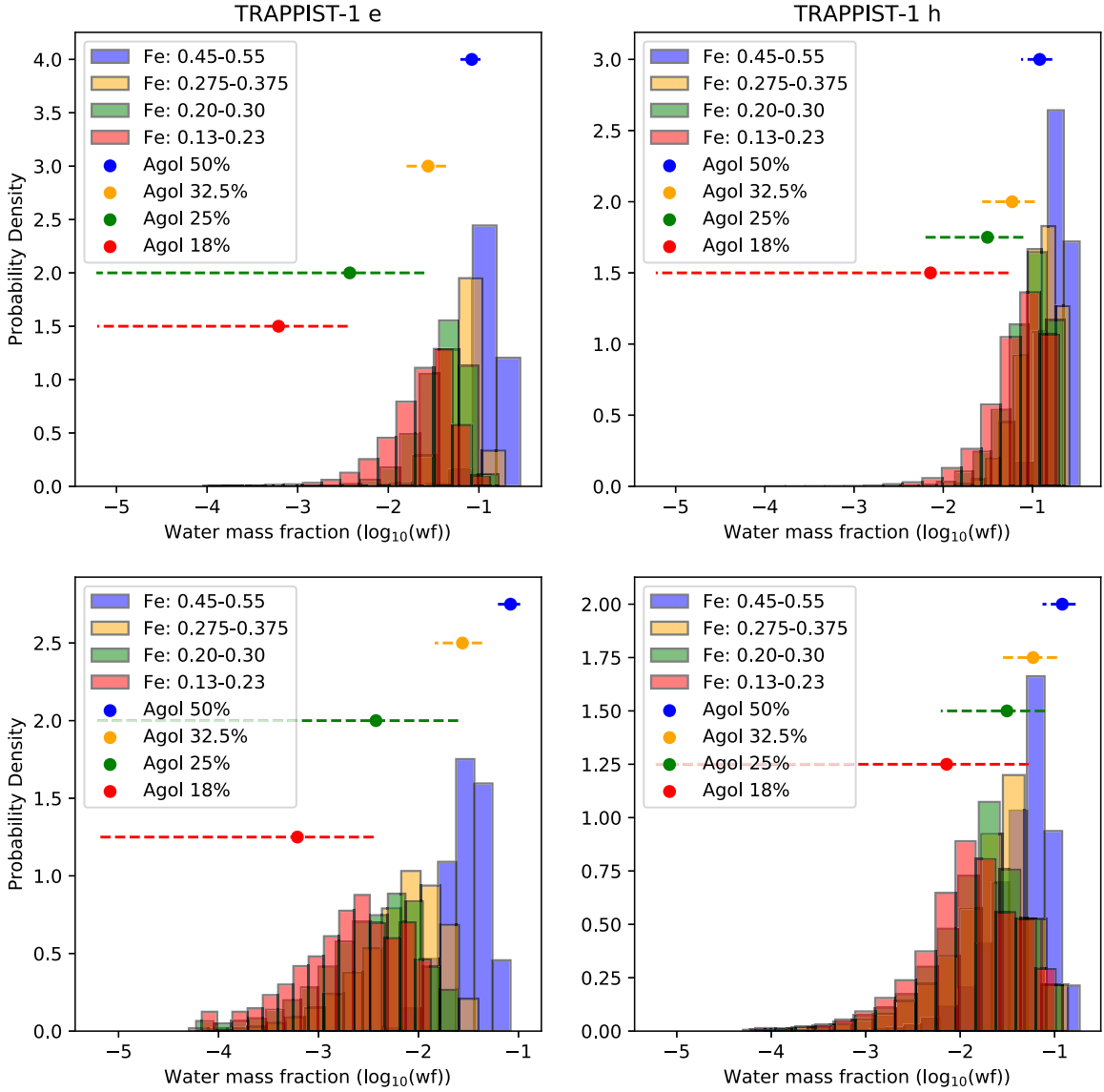

**Figure B2.** Identical to Figure A2 except calculations have been repeated for Trappist-1e (left) and Trappist-1h (right).

## Appendix C
## Atmospheric Evolution Model

### C.1. Seafloor Weathering Parameterization

The coupled atmosphere–interior model used to generate Figure 4 in the main text is described in full in Krissansen-Totton et al. (2021), and the code is available at https://github.com/joshuakt/Oxygen-False-Positives. This section describes the modifications made to this model to simulate Trappist-1 waterworlds.

The seafloor weathering flux is parameterized as follows:

$$F_{\text{seafloor weathering}} = \min \{F_{\text{kinetic}}, F_{\text{supply limit}}\}$$
$$F_{\text{supply limit}} = [\text{DIC}_{\text{ocean}}] M_{CO_2} F_{\text{hydro}}. \quad (C1)$$

Here, seafloor weathering is taken to be the minimum of the kinetics-dependent dissolution of cations, and the supply of carbon in hydrothermal fluids. The supply limit to seafloor weathering, $F_{\text{supply limit}}$, is set by the amount of dissolved inorganic carbon, $[\text{DIC}_{\text{ocean}}]$, that is delivered to oceanic crust via hydrothermal fluxes, $F_{\text{hydro}}$ (kg yr$^{-1}$), which in turn is assumed to depend linearly on interior heatflow (Coogan & Gillis 2013).

A nominal range for the modern Earth's hydrothermal flux, $F_{\text{hydro}}(\text{heatflow} = 1)$, is sampled uniformly from $10^{15}$ kg yr$^{-1}$ to $6 \times 10^{16}$ kg yr$^{-1}$ (Krissansen-Totton & Catling 2017). Here, $M_{CO_2}$, is the molar mass of carbon dioxide. The rate at which cations are dissolved from the seafloor, $F_{\text{kinetic}}$, is controlled by the temperature ($T_{\text{deep}}$) and pH ($pH$) of hydrothermal fluids, as well as plate velocity, $\nu_{\text{plate}}$:

$$F_{\text{kinetic}} = \frac{W_{\text{coef}}}{4} \times 10^{(-0.3(\text{pH}-7.7))} \left(\frac{\nu_{\text{plate}}}{\nu_{\text{plate}}^{\text{Earth}}}\right)$$
$$\times \exp\left(-\frac{E_{\text{SF}}}{8.314 \text{ J mol}^{-1} \text{ K}^{-1}} \left(\frac{1}{T_{\text{deep}}} - \frac{1}{285 \text{ K}}\right)\right) \quad (C2)$$

Here, $W_{\text{coef}} = 1.3 \times 10^{11}$ kg yr$^{-1}$ is a constant scaled to fit Earth fluxes, $\nu_{\text{plate}}^{\text{Earth}} = 0.03$ m yr$^{-1}$ is the average plate speed on the modern Earth, and $E_{\text{SF}} = 90$ kJ mol$^{-1}$ is an effective activation energy. Plate speed, $\nu_{\text{plate}}$, is calculated assuming the following relationship between melt production, MP (m$^3$ yr$^{-1}$),

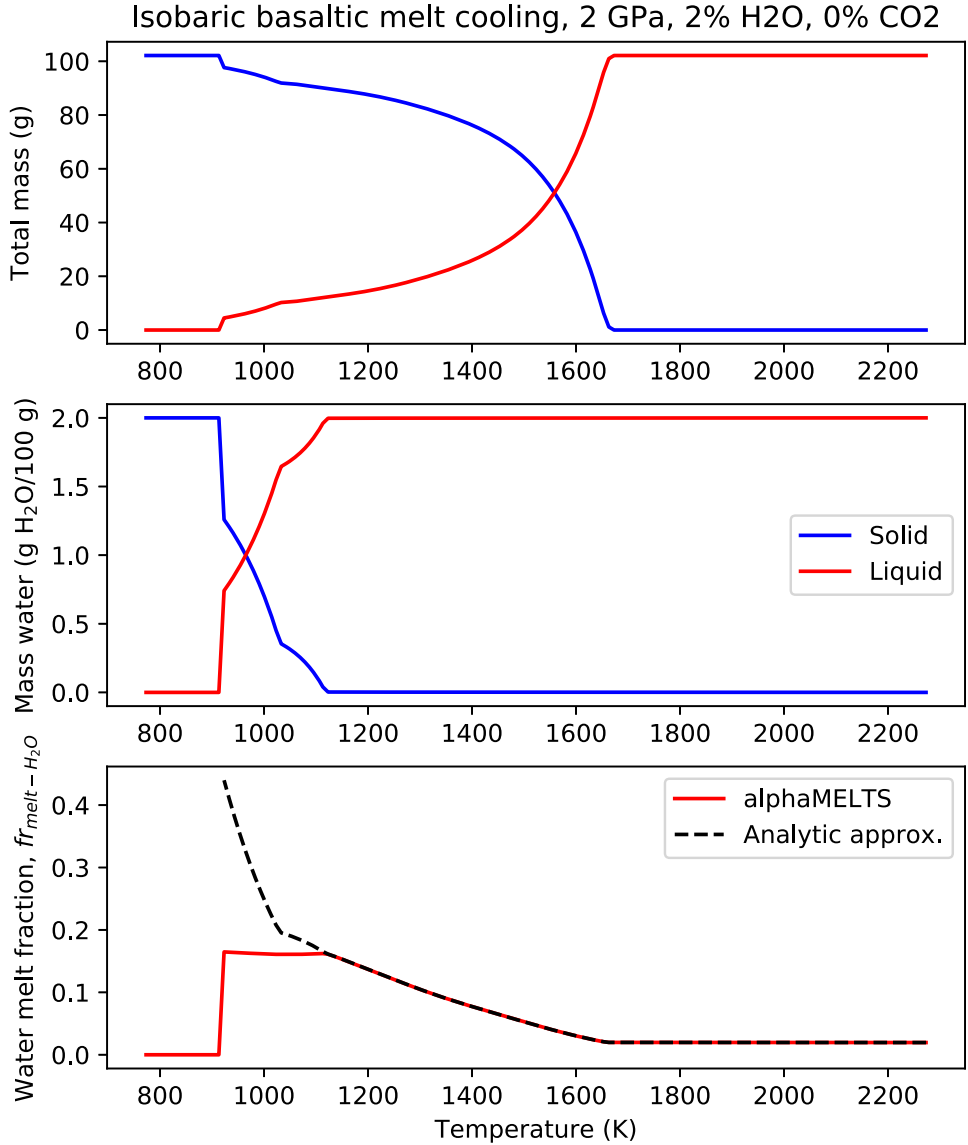

**Figure C1.** Example calculations from thermodynamic melt crystallization model alphaMELTS showing the constant-pressure cooling of a 100 g basaltic melt. (a) Total mass of crystalline solids (blue) and liquids (red), (b) mass of water dissolved in the melt (red) and water incorporated into crystalline solids (blue), and (c) water mass fraction dissolved in the remaining melt (red). The black-dashed line shows the analytic approximation to water melt fraction (Equation (C4)) used in our planetary evolution model.

ridge length, $l_{\rm ridge}$ (m), and crustal depth, $d_{\rm crust}$ (m):

$$\mathrm{MP} = l_{\rm ridge} \times d_{\rm crust} \times v_{\rm plate}. \qquad (C3)$$

This reduces to an overall dependence of plate velocity on interior heatflow of $v_{\rm plate} \propto \sqrt{\rm heatflow}$ (Krissansen-Totton et al. 2021). Here, we are assuming that only $CO_2$ availability limits seafloor weathering, and that the supply of weatherable oceanic, represented by plate velocity, scales freely with heatflow.

### C.2. Melt Fraction and Volatile Exsolution

In the planetary evolution model used to produce Figure 4, the concentrations of water and carbon dissolved in partial melts are given by

$$fr_{\rm melt\text{-}H_2O} = \frac{(1 - (1-\bar{\psi})^{1/k_{\rm H_2O}})}{\bar{\psi}} \frac{M_{\rm solid\text{-}H_2O}}{M_{\rm mantle}} \qquad (C4)$$

$$fr_{\rm melt\text{-}CO_2} = \frac{(1 - (1-\bar{\psi})^{1/k_{\rm CO_2}})}{\bar{\psi}} \frac{M_{\rm solid\text{-}CO_2}}{M_{\rm mantle}}. \qquad (C5)$$

Here, $k_{\rm H_2O} = 0.01$ and $k_{\rm CO_2} = 2 \times 10^{-3}$ are the constant partition coefficients assumed for melt–solid partitioning of volatiles, $M_{\rm solid\text{-}H_2O}/M_{\rm mantle}$ is the concentration of water in mantle–source material, $M_{\rm solid\text{-}CO_2}/M_{\rm mantle}$ is the concentration of carbon dioxide in mantle–source material, and $\bar{\psi}$ is the average melt fraction over the portion of the mantle where melting occurs (determined by the interior evolution model). Here, we compare this simplified approach to estimating volatile melt content to more detailed thermodynamic calculations using alphaMELTS (Ghiorso et al. 2002; Smith & Asimow 2005). Figure C1 shows example calculations from alphaMELTS illustrating the constant-pressure cooling of a basaltic melt at 2 GPa with 2 wt% water. Figure C1(a) shows sold and liquid mass fractions as a function of temperature,

Figure C1(b) shows the total mass of water dissolved in the melt alongside solid phase water, and Figure C1(c) shows the water melt fraction from alphaMELTs calculations alongside the simplified expression given by Equation (C4). The close match confirms that our planetary evolution model reasonably approximates volatile concentration in partial melts at high pressures. The two estimates diverge at low melt fractions because alphaMELTs predicts the precipitation of hydrated minerals, contrary to the $k_{H_2O} = 0.01$ approximation. The comparatively high melt volatile concentrations also illustrate why gravitational segregation of intrusive melts could lead to crust–ocean volatile exchange (see Section 4).

### ORCID iDs

Nicholas Wogan https://orcid.org/0000-0002-0413-3308
Jonathan J. Fortney https://orcid.org/0000-0002-9843-4354

### References

Agol, E., Dorn, C., Grimm, S. L., et al. 2021, PSJ, 2, 1
Anderson, K. R., & Poland, M. P. 2017, NatGe, 10, 704
Barnes, R., Mullins, K., Goldblatt, C., et al. 2013, AsBio, 13, 225
Barth, P., Carone, L., Barnes, R., et al. 2020, arXiv:2008.09599
Bottinga, Y., & Javoy, M. 1990, JGR, 95, 5125
Burgisser, A., & Degruyter, W. 2015, Magma Ascent and Degassing at Shallow Levels The Encyclopedia of Volcanoes (Amsterdam: Elsevier), 225
Busigny, V., Laverne, C., & Bonifacie, M. 2005, GGG, 6, Q12O01
Choblet, G., Tobie, G., Sotin, C., Kalousova, K,, & Grasset, O. 2017, Icar, 285, 252
Coogan, L. A., & Gillis, K. M. 2013, GGG, 14, 1771
Coogan, L. A., & Gillis, K. M. 2018, AREPS, 46, 21
Cowan, N. B., & Abbot, D. S. 2014, ApJ, 781, 27
De Wit, J., Wakeford, H. R., Lewis, N. K., et al. 2018, NatAs, 2, 214
Dixon, J. E. 1997, AmMin, 82, 368
Dorn, C., Noack, L., Rozel, A., et al. 2018, A&A, 614, A18
Dziewonski, A. M., & Anderson, D. L. 1981, PEPI, 25, 297
Edmonds, M., & Woods, A. W. 2018, JVGR, 368, 13
Elkins-Tanton, L. T. 2008, E&PSL, 271, 181
Esposito, R., Hunter, J., Schiffbauer, J. D., Shimizu, N., & Bodnar, R. J. 2014, AmMin, 99, 976
Fauchez, T. J., Turbet, M., Villanueva, G. L., et al. 2019, ApJ, 887, 194
Fischer, T. P., & Marty, B. 2005, JVGR, 140, 205
Foley, B. J., & Smye, A. J. 2018, AsBio, 18, 873
Gaetani, G. A., O'Leary, J. A., Shimizu, N., Bulcholz, C. E., & Newville, M. 2012, Geo, 40, 915
Gaillard, F., & Scaillet, B. 2014, E&PSL, 403, 307
Gale, A., Dalton, C. A., Langmuir, C. H., Su, Y., & Schilling, J. G. 2013, GGG, 14, 489
Garcia, M. O., Frey, F. A., Grooms, D. G., et al. 1986, CoMP, 94, 461
Ghiorso, M. S., & Gualda, G. A. 2015, CoMP, 169, 1
Ghiorso, M. S., Hirschmann, M. M., Reiners, P. W., & Kress, V. C. 2002, GGG, 3, 1
Glaser, D. M., Hartnett, H. E., Desch, S. J., et al. 2020, ApJ, 893, 163
Grimm, S. L., Demory, B.-O., Gillon, M., et al. 2018, A&A, 613, A68
Hamano, K., Abe, Y., & Genda, H. 2013, Natur, 497, 607
Hayworth, B. P., & Foley, B. J. 2020, ApJL, 902, L10
Holloway, J. R., & Blank, J. G. 1994, Reviews in Mineralogy, 30, 187
Höning, D., Tosi, N., Spohn, T., et al. 2019, A&A, 627, A48
Iacono-Marziano, G., Morizet, Y., Le Trong, E., & Gaillard, F. 2012, GeCoA, 97, 1
Ikoma, M., Elkins-Tanton, L., Hamano, K., & Suckale, J. 2018, SSRv, 214, 76
Journaux, B., Daniel, I., Petitgirard, S., et al. 2017, E&PSL, 463, 36
Journaux, B., Kalousová, K., Sotin, C., et al. 2020, SSRv, 216, 7
Kalousová, K., Sotin, C., Choblet, G., Tobie, G., & Grasset, O. 2018, Icar, 299, 133
Katz, R. F., & Cashman, K. V. 2003, GGG, 4, 8705
Kerrick, D., & Connolly, J. 2001, E&PSL, 189, 19
Kite, E. S., & Ford, E. B. 2018, ApJ, 864, 75
Kite, E. S., Manga, M., & Gaidos, E. 2009, ApJ, 700, 1732
Kitzmann, D., Alibert, Y., Godolt, M., et al. 2015, MNRAS, 452, 3752
Krissansen-Totton, J., & Catling, D. C. 2017, NatCo, 8, 15423
Krissansen-Totton, J., Fortney, J. J., Nimmo, F., & Wogan, N. 2021, AGUA, 2, e2020AV000294
Krissansen-Totton, J., Garland, R., Irwin, P., & Catling, D. C. 2018, AJ, 156, 114
Le Voyer, M., Hauri, E. H., Cottrell, E., et al. 2019, GGG, 20, 1387
Libourel, G., Marty, B., & Humbert 2003, GeCoA, 67, 4123
Luger, R., & Barnes, R. 2015, AsBio, 15, 119
Lustig-Yaeger, J., Meadows, V. S., & Lincowski, A. P. 2019, AJ, 158, 27
Manning, C. E., Shock, E. L., Sverjensky, D. A., et al. 2013, RvMG, 75, 109
Marrocchi, Y., & Toplis, M. 2005, GeCoA, 69, 5765
Marty, B. 2012, E&PSL, 313, 56
Matyska, C., & Yuen, D. A. 2001, E&PSL, 189, 165
Moore, G., Vennemann, T., & Carmichael, I. 1998, AmMin, 83, 36
Moore, L. R., Gazel, E., Tuohy, R., et al. 2015, AmMin, 100, 806
Nakayama, A., Kodama, T., Ikoma, M., & Abe, Y. 2019, MNRAS, 488, 1580
Newman, S., & Lowenstern, J. B. 2002, CG, 28, 597
Nisr, C., Chen, H., Leinenweber, K., et al. 2020, PNAS, 117, 9747
Noack, L., Höning, D., Rivoldini, A., et al. 2016, Icar, 277, 215
Noack, L., Rivoldini, A., van Hoolst, T., et al. 2017, PEPI, 269, 40
Ortenzi, G., Noack, L., Sohl, F., et al. 2020, NatSR, 10, 1
Pasek, M., Omran, A., Lang, C., et al. 2020, Serpentinization as a Route to Liberating Phosphorus on Habitable Worlds. doi:10.21203/rs.3.rs-37651/v1
Quick, L. C., Roberge, A., Mlinar, A. B., & Hedman, M. M. 2020, PASP, 132, 084402
Raymond, S. N., Kokubo, E., Morbidelli, A., Morishima, R., & Walsh, K. J. 2013, arXiv:1312.1689
Raymond, S. N., Quinn, T., Lunine, J. I., et al. 2004, Icar, 168, 1
Roskosz, M., Bouhifd, M. A., Jephcoat, A., Marty, B., & Mysen, B. 2013, GeCoA, 121, 15
Schaefer, L., Wordsworth, R. D., Berta-Thompson, Z., & Sasselov, D. 2016, ApJ, 829, 63
Seager, S., Kuchner, M., Hier-Majumder, C., & Militzer, B. 2007, ApJ, 669, 1279
Smith, P. M., & Asimow, P. D. 2005, GGG, 6, Q02004
Staudigel, H. 2003, TrGeo, 3, 659
Syverson, D. D., Reinhard, C. T., Isson, T. T., et al. 2020, arXiv:2002.07667
Tajika, E., & Matsui, T. 1992, E&PSL, 113, 251
Tosi, N., Godolt, M., Stracke, B., et al. 2017, A&A, 605, A71
Turbet, M., Ehrenreich, D., Lovis, C., Bolmont, E., & Fauchez, T. 2019, A&A, 628, A12
Tutolo, B. M., Seyfried, W. E., & Tosca, N. J. 2020, PNAS, 117, 14756
Unterborn, C. T., Desch, S. J., Hinkel, N. R., & Lorenzo, A. 2018, NatAs, 2, 297
Unterborn, C. T., Kabbes, J. E., Pigott, J. S., Reaman, D. M., & Panero, W. R. 2014, ApJ, 793, 124
Vazan, A., Sari, R. E., & Kessel, R. 2020, arXiv:2011.00602
Wallace, P. J., Plank, T., Edmonds, M., & Hauri, E. H. 2015, Volatiles in Magmas The Encyclopedia of Volcanoes (Amsterdam: Elsevier), 163
Wogan, N., Krissanson-Totton, J., & Catling, D. C. 2020, PSJ, 1, 58
Wordsworth, R., Schaefer, L., Fischer, R., et al. 2018, AJ, 155, 195
Wunderlich, F., Godolt, M., Grenfell, J. L., et al. 2019, A&A, 624, A49
Zeng, L., Jacobsen, S. B., Sasselov, D. D., et al. 2019, PNAS, 116, 9723
Zeng, L., & Seager, S. 2008, PASP, 120, 983